\documentclass[aps,prl,reprint,superscriptaddress,amsmath,amssymb,floatfix,showkeys]{revtex4-2}
\usepackage{graphicx}% Include figure files
\usepackage{dcolumn}% Align table columns on decimal point
\usepackage{bm}% bold math
\usepackage{color}
\begin{document}
%%%%%%%%%%% Título, autores y resumen %%%%%%%%%
\title{Determining depletion interactions by contracting forces}
\author{N\'estor M.\ de los Santos-L\'opez}
%\email{nestor.santos@cinvestav.mx}
\affiliation{Departamento de F\'isica Aplicada, Cinvestav-M\'erida, AP 73 ``Cordemex'', 97310 M\'erida, Yucat\'an, Mexico.}
\author{Gabriel P\'erez-\'Angel}
\email{gperez@cinvestav.mx}
\affiliation{Departamento de F\'isica Aplicada, Cinvestav-M\'erida, AP 73 ``Cordemex'', 97310 M\'erida, Yucat\'an, Mexico.}
\author{Ram\'on Casta\~neda-Priego}
%\email{ramoncp@fisica.ugto.mx}
\affiliation{Divisi\'on de Ciencias e Ingenier\'ias, Campus Le\'on, Universidad de Guanajuato, Loma del Bosque 103, 37150 Le\'on, Guanajuato, Mexico.}
\author{José M. M\'endez-Alcaraz}
%\email{jmendez@fis.cinvestav.mx}
\affiliation{Departamento de F\'isica, Cinvestav, Av.\ IPN 2508, Col.\ San Pedro Zacatenco, 07360 Gustavo A.\ Madero, Ciudad de M\'exico, Mexico.}
\date{\today}
\begin{abstract}
We introduce a general physical formulation that allows one to obtain uniquely the effective interactions between particles by contracting the bare forces, even in highly concentrated systems. We tested it by studying depletion forces in binary and ternary colloidal mixtures with a total packing fraction up to 55\%. Our result opens up the possibility of finding an efficient route to determine effective interactions at finite concentration and even at thermodynamic conditions near to meta-stable or out of equilibrium states.

\end{abstract}
\maketitle
%%%%%%%%%% Texto de la carta %%%%%%%%%%
The universal mechanism behind depletion forces in colloidal-like systems with large size-asymmetries and repulsive short-ranged interactions (predominantly in those systems whose main bare interaction is of the hard-core type) consists in the spontaneous expulsion of small particles from the gap between two approaching large particles, leading to an imbalance of osmotic pressures on their internal and external faces, which drive them to attract each other giving rise to a decrease in the free energy of the system that results in a strong, effective attraction at contact~\cite{AO1,lekkerkerker}.

When analyzing the phenomenon of depletion forces beyond the dilute limit, one typically finds that a repulsive wall just after the attractive well at contact emerges \cite{alcaraz-klein,ramon-ro-al-2,Roth2000,Roth2011}. This effective repulsion is due to the formation of a layer of first neighbors of small particles around the large ones, when they are separated by a distance larger than the diameter of the former. This leads to an overpopulation of small particles in the gap between large ones, exactly the opposite of the depletion at smaller separations, driving again to an unbalance of osmotic pressures but now pushing the larger particles out. This barrier works as a kind of gate that controls the action of the attractive well at contact~\cite{Nestor2021}, and hence its enormous relevance for understanding, for example, the formation of colloidal clusters or the thermodynamic stability of the dispersion~\cite{Nestor2018}.

Theoretical, experimental, and simulation studies made it possible to understand in detail the behavior of depletion forces as a function of the concentration, size, and morphology of the involved species~\cite{lekkerkerker}. Depletion forces have been identified as the physical mechanism behind important thermodynamic processes and biological phenomena~\cite{lekkerkerker}. They have also led the development of entropic engineering techniques~\cite{dinsmore,lin_PRL_2000}. Currently, there has been a growing interest to understand the nature of depletion forces under non-equilibrium conditions~\cite{Dzubiella2003} and near thermodynamic meta-stable states~\cite{erik,Lu2008,Gnan2014}.

There exist several theoretical approaches able to capture the complexity of depletion forces~\cite{lekkerkerker,alcaraz-klein,ramon-ro-al-2,Roth2000,Roth2011}. Among them, we refer below to the integral equations theory~\cite{alcaraz-klein,ramon-ro-al-2} to set a context of our results and help in the discussion of their physical interpretation. From the perspective of this formalism, the structure of those particles that belong to the same species in a mixture can be fully described by an effective potential that results from the integration of the degrees of freedom of the remaining species, leading to the same distribution that would be obtained from a complete description in terms of the bare potentials among all species. An interesting and natural result obtained from the integral equations theory of depletion forces is that the effective potential between large particles immersed in a bath of small ones tends to the potential of mean force when only a few of the former are present, even for high concentrations of the latter. This prediction has been confirmed by independent experiments and molecular simulations \cite{erik,perera}.

The direct comparison between theoretical effective interactions and computer simulations has only been possible in the diluted case of large particles, in which depletion forces can be obtained from the simulations by fixing two large particles at a given separation and adding the projections on the line connecting their centers of the transfer of momentum per unit time due to the collisions between them and the small particles~\cite{Biben1996,ramon-ro-al-2}. This is, in fact, the standard protocol to extract an effective potential from molecular dynamics simulations even in those cases where the small particles undergo a phase transition or are near a percolated-like state~\cite{Gnan2014}. Up to now, a comparison for high concentrations of large particles has not been possible due to the lack of a physical approach able to extract the effective interactions from the simulations under those conditions. Thus, this Letter aims to introduce a general physical formulation that allows us to obtain the depletion forces even when the concentration of the non-depleted is not negligible.

In the following, we will show that the effective interactions, particularly the depletion one, can be uniquely determined from the contraction of the forces. Let us assume a colloidal mixture composed of $N_l$ large spheres and $N_s$ small ones, generically denoted with the sub-index $s$ for the moment, but leaving open the possibility of several small species. The total force exerted on the $i$-th large particle at time $t$ is
\begin{equation}
\bm{F}^l_i (\bm{r}_i ) = \sum_{j \neq i}^{N_l} \bm{f}^{ll}_{ji}(\bm{r}_{ji}) +
\sum_{k=1}^{N_s} \bm{f}^{sl}_{ki}(\bm{r}_{ki}),
\label{ftot}
\end{equation}
where $\bm{f}^{ll}_{ji}(\bm{r}_{ji})$ is the bare force exerted by the $j$-th large particle and $\bm{f}^{sl}_{ki}(\bm{r}_{ki})$ the one exerted by the $k$-th small particle. Now, in the contracted description, we rewrite equation (\ref{ftot}) in the form
\begin{equation}
\langle \bm{F}^l_i (\bm{r}_i ) \rangle_s = \sum_{j \neq i}^{N_l} \bm{G}^{ll}_{ji}(\bm{r}_{ji}),
\label{ftotcont}
\end{equation}
where $\langle \bm{F}^l_i (\bm{r}_i ) \rangle_s$ is the average of $\bm{F}^l_i (\bm{r}_i )$ over the configurations of the small particles in the fixed field of the large ones. Therefore, $u^\mathrm{eff}_{ll}(r_{ji})$ in  $\bm{G}^{ll}_{ji}(\bm{r}_{ji})=G^{ll}_{ji}(r_{ji}) \hat{\bm{r}}_{ji}=- \bm{\nabla} u^\mathrm{eff}_{ll}(r_{ji})$ shall correspond to the theoretical effective interaction potential between large particles. Here, the time dependence enters into the equations through the particle positions; $\bm{r}_i$ is the position of the $i$-th particle and $\bm{r}_{ji}=\bm{r}_{i}-\bm{r}_{j}$ the position of particle $i$ respect to particle $j$, being $r_{ji}$ the distance between their centers at time $t$. Also $\hat{\bm{r}}_{ji}$ is a unitary vector along $\bm{r}_{ji}$, and $\bm{\nabla} = d/d\bm{r}_{ji}=\hat{\bm{r}}_{ji} d/dr_{ji}$.

Small particles were integrated out from the description by going from Eqs.~(\ref{ftot}) to (\ref{ftotcont}), as they go from being explicit components of the system to a part of the supporting environment, after averaging over their configurations. Their effects are implicitly included in the effective force $\bm{G}^{ll}_{ji}(\bm{r}_{ji})$ between large particles. Our idea is to evaluate $\bm{F}^l_i (\bm{r}_i )$ using Eq.~(\ref{ftot}) and then take it into Eq.~(\ref{ftotcont}) to get $\bm{G}^{ll}_{ji}(\bm{r}_{ji})$ through an inversion process. The instantaneous values of $\bm{F}^l_i (\bm{r}_i )$ obtained from Eq.~(\ref{ftot}) are noisy, even if they are obtained through a deterministic simulation run, and satisfy Eq.~(\ref{ftotcont}) only over long times. Instead, they may be described by the stochastic equation,
\begin{equation}
\bm{F}^l_i (\bm{r}_i ) = \sum_{j \neq i}^{N_l} \bm{G}^{ll}_{ji}(\bm{r}_{ji})+\bm{D}^{l}_{i}(t),
\label{ftotcont2}
\end{equation}
where $\bm{D}^{l}_{i}(t)$ corresponds to the non-systematic part of the instant forces, due to the constant tapping of small particles on the large ones, expected to be a noisy term with zero mean, $\langle \bm{D}^l_i (t) \rangle_s =0$, and finite variance (irrelevant for the determination of the effective interaction). Furthermore, the force $\bm{F}^l_i (\bm{r}_i )$ in Eq.~(\ref{ftotcont2}) can be interpreted as the noisy result of a measurement and the $\bm{G}^{ll}_{ji}(\bm{r}_{ji})$'s as fitting functions for it, being $\bm{D}^{l}_{i}(t)$ the instant deviations between measured and fitted values. Therefore, the values of $\bm{G}^{ll}_{ji}(\bm{r}_{ji})$ can be obtained using the least-squares method~\cite{numerical-recipes}.

To illustrate the applicability of the contraction of the forces given by Eq. (\ref{ftotcont2}), we now study the depletion forces in colloidal mixtures. After getting $\bm{F}^l_i (\bm{r}_i )$ from Eq.~(\ref{ftot}), we split Eq.~(\ref{ftotcont2}) into its Cartesian components,
\begin{equation}
F_{i,\alpha}^l (\bm{r}_i ) = \sum_{j\neq i}^{N_l} G^{ll}_{ji} (r_{ji}) \cos \theta_{ji,\alpha} + D_{i,\alpha}^l (t),
\label{fsplited}
\end{equation}
with $\alpha=x,y,z$ and $\cos \theta_{ji,\alpha}$ the corresponding directional cosines. To proceed numerically, we discretize the distance $r_{ji}$ into $C$ classes,
\begin{equation}
r_{ji} \rightarrow r_n = n \delta, \quad \mbox{if} \quad r_n \le r_{ji} < r_{n+1},
\label{clases}
\end{equation}
where $\delta$ is the class-size and $n = 0,...,C-1$, assuming $r_{max} = C \delta$ as the range of the depletion effects. Therefore, the forces become
\begin{equation}
F_{i,\alpha}^l (\bm{r}_i ) = \sum_{n=0}^{C-1} A_{n,\alpha}^i G^{ll}_{ji} (r_n) + D_{i,\alpha}^l (t),
\label{fsplited2}
\end{equation}
with
\begin{equation}
A_{n,\alpha}^i = \sum_{j\neq i}^{N_{l,n}} \cos \theta_{ji,\alpha}.
\label{fsplited3}
\end{equation}
The sub-index $n$ in the upper limits of the sums indicates that these are carried out only over those $j$-values with $r_{ij}$ being in class $n$. In matrix notation, one gets
\begin{equation}
\mathbb{F}_\alpha = \mathbb{A}_\alpha \mathbb{G} + \mathbb{D}_\alpha.
\label{matrix}
\end{equation}
The matrices $\mathbb{F}_\alpha$ and $\mathbb{D}_\alpha$ have only one column with $\mathcal{N} N_l$ rows, being $\mathcal{N}$ the number of simulated configurations used to gather statistics, with typical values for this product around $10^{6}$ or more, since one accumulates in the matrix arrangements the values obtained for each of the analyzed configurations, for each large particle. The matrix $\mathbb{G}$ has one column with $C$ rows, typically about $10^{2}$. The matrices $\mathbb{A}_\alpha$ are arrays with the same number of rows as $\mathbb{F}_\alpha$, and $C$ columns. 

The linear equations represented by (\ref{matrix}) are highly over-determined, since the rows are much more than the columns. To overcome this difficulty, we use the least-squares method (LS)~\cite{numerical-recipes}, demanding the values of $\mathbb{G}$'s elements to minimize the ones of $\mathbb{D}^\mathtt{T}_\alpha \mathbb{D}_\alpha$. This leads to equation
\begin{equation}
\mathbb{A}^\mathtt{T}_\alpha \mathbb{A}_\alpha \mathbb{G} = \mathbb{A}^\mathtt{T}_\alpha \mathbb{F}_\alpha,
\label{matrix3}
\end{equation}
which represent closed and well-defined systems of $C$ linear equations with $C$ variables, which are solved using Singular Value Decomposition (SVD), since the matrices
$\mathbb{A}^\mathtt{T}_\alpha \mathbb{A}_\alpha$ may be singular or close to it~\cite{numerical-recipes}. We carry out the same procedure for all $\mathbb{F}_\alpha$, with $\alpha=x,y,z$, independently, and so one gets three different solutions for $\mathbb{G}$. Clearly, a comparison between them gives some validation to our approach; in all cases, we found them equal up to statistical fluctuations. The values shown below are the average of all solutions. At this point, we should set the details of the system under consideration.

We simulate ternary mixtures composed of large, medium, and small particles, denoted as $l$, $m$, and $s$, with diameters $\sigma_{l}$, $\sigma_{m} = \sigma_{l}/3$, and $\sigma_{s} =\sigma_{l}/5$, respectively. The bare interaction between them is given by the Mie 49-50 potential,
\begin{eqnarray}
u_{ij} (r_{ij}) & = &
A \epsilon \left[ \left( \frac{\sigma_{ij}}{r_{ij}}\right)^{50} -\left( \frac{\sigma_{ij}}{r_{ij}}\right)^{49} \right] + \epsilon \;\; \mbox{if} \;\; r_{ij} < \sigma_{ij} B \nonumber \\
& = & 0 \;\; \mbox{otherwise},
\label{WCA}
\end{eqnarray}
whith $\sigma_{ij} = (\sigma_i + \sigma_j)/2$, $B=50/49$, $A=50B^{49}$, and $\beta \epsilon=0.6785$, which provides a reasonable approach to hard particles, since it matches their second virial coefficient~\cite{Baez2018}. We perform Molecular Dynamics (MD) simulations in a cubic box of volume $V$ with periodic boundary conditions in the canonical ensemble at a reduced temperature $T^{*}\equiv k_{B} T/\epsilon = 1.0$. Length, mass, and time scales are given by $\sigma_l$, $m_l$, and $\sqrt{m_l \sigma_l^2 / k_B T}$, respectively. The mass of a large particle is $m_l$, while the masses of the smaller ones keep the same proportion as their volumes. The total number of particles is $N=N_l +N_m+N_s$, and $\phi_i = \pi \sigma_i^3 N_i / 6 V$ is the volume fraction of the $i$-th species, with $\phi=\phi_l +\phi_m+\phi_s$ being the total packing fraction. Typical values in our calculations are $V^{1/3}=10 \sigma_l$ and $N=16384$. Here, $\beta=1/k_B T$ is the inverse of the thermal energy, with $k_{B}$ being the Boltzmann constant and $T$ the absolute temperature.

The simulation begins by randomly placing the particles without overlapping and it is carried out using the velocity-Verlet algorithm~\cite{Allen1987}. To keep the temperature constant, a simple velocity rescaling criterion is used. The transient part is run until equilibrium is reached, checked by monitoring the excess energy, and waiting until it becomes stationary; to avoid possible aging effects, it is ensured that the mean-square displacements and radial distribution functions (RDFs) do not depend on the transient time in systems with $\phi$ above 45\%. We have also verified that the initial and final RDFs for the production runs are alike. Afterwards, the calculation of the forces starts. Applying the test of self-consistency, we also simulate with MD the effective monodisperse systems of large particles interacting via the numerically extracted depletion forces. The RDFs for those monodisperse runs are then compared with their counterparts in the original simulations for the ternary mixture with the bare interactions.

\begin{figure}[htbp!]
\includegraphics[scale=0.42]{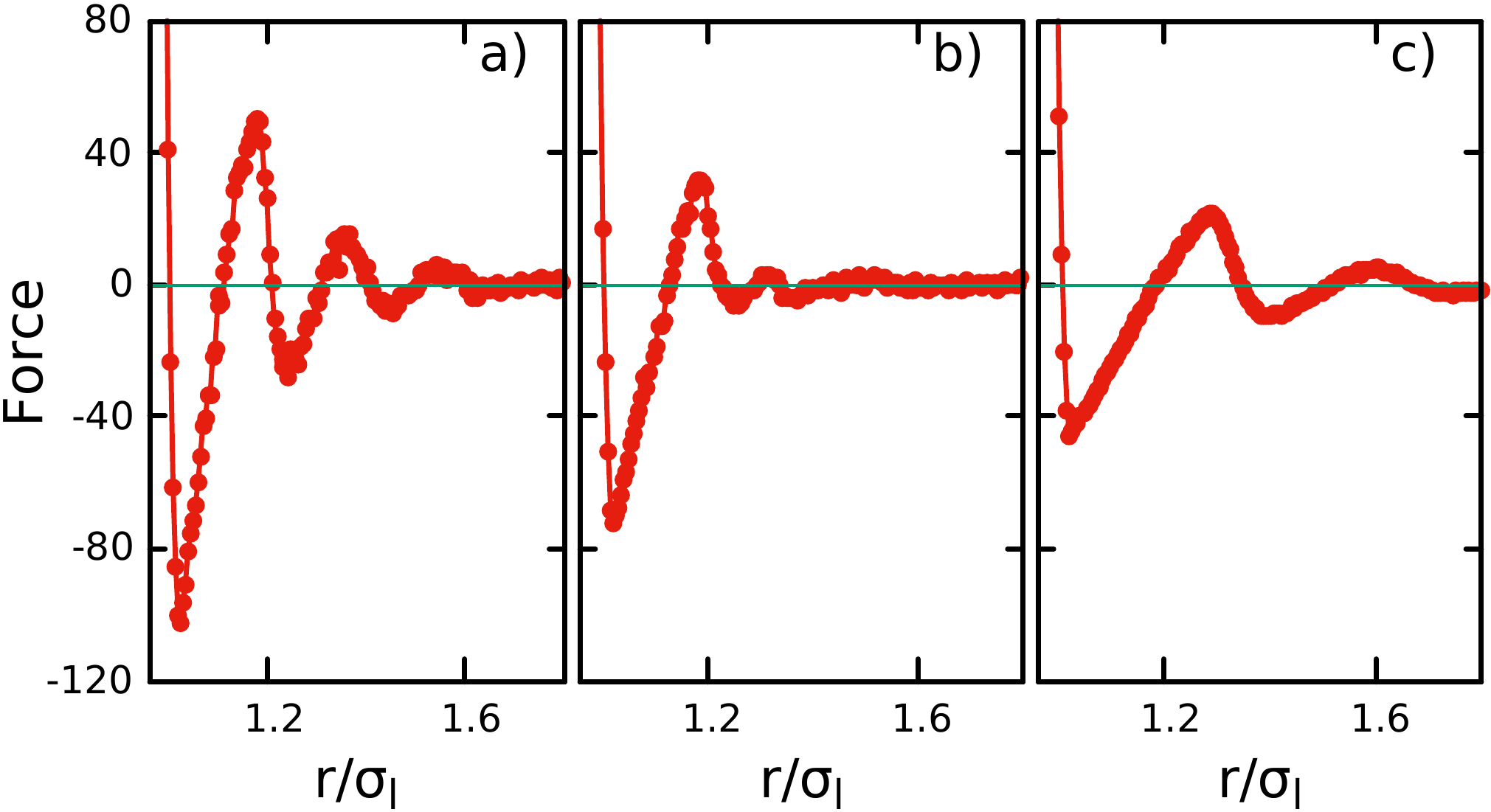}
\caption{Depletion force, times $\beta \sigma_{l}$, for three systems with $\phi=0.45$, having $\phi_{l}=0.15$ and: (a) $\phi_m=0$ and $\phi_s=0.3$; (b) $\phi_m=\phi_s=0.15$; (c) $\phi_m=0.3$ and $\phi_s=0.0$.}
\label{fig1}
\end{figure}

Figure \ref{fig1} shows the depletion force, times $\beta \sigma_{l}$, as a function of the distance between the centers of large particles for three systems with $\phi=0.45$ and $\phi_{l}=0.15$ fixed. The extreme repulsion about contact comes from the bare potential and ensures volume exclusion. The following attractive well corresponds to the depletion attraction that can be understood quite well in terms of the Asakura-Oosawa approximation~\cite{AO1,alcaraz-klein,ramon-ro-al-2,Nestor2021}. In the bidisperse cases, its depth is proportional to the concentration of depletants and the size ratio $\sigma_l /\sigma_s$, while its range is the size of the depletants. It is followed by the repulsive barrier arising from the accumulation of depletants in the first neighbors layer around large particles, a complex and poorly studied effect ~\cite{ramon-ro-al-2}; it is, however, of importance for the understanding of cluster formation in competing interaction fluids~\cite{Douglas2014,Nestor2021}. At first glance, its amplitude and range seem to behave like those of the attractive well, in the bidisperse cases. Nevertheless, although not shown here, the main consequence of adding a second species of depletants is to modify this barrier in such a way that polydispersity seems to be a control parameter of this entropic gate~\cite{Roth2006,Nestor2021}. A similar outcome has been observed in binary mixtures of large spheres and small spherocylinders of diameter $\sigma$ and length $L$, which roughly behave like two spherical depletant species of diameters $\sigma$ and $L$~\cite{Pedro2005}.

\begin{figure}[htbp!]
\includegraphics[scale=0.55]{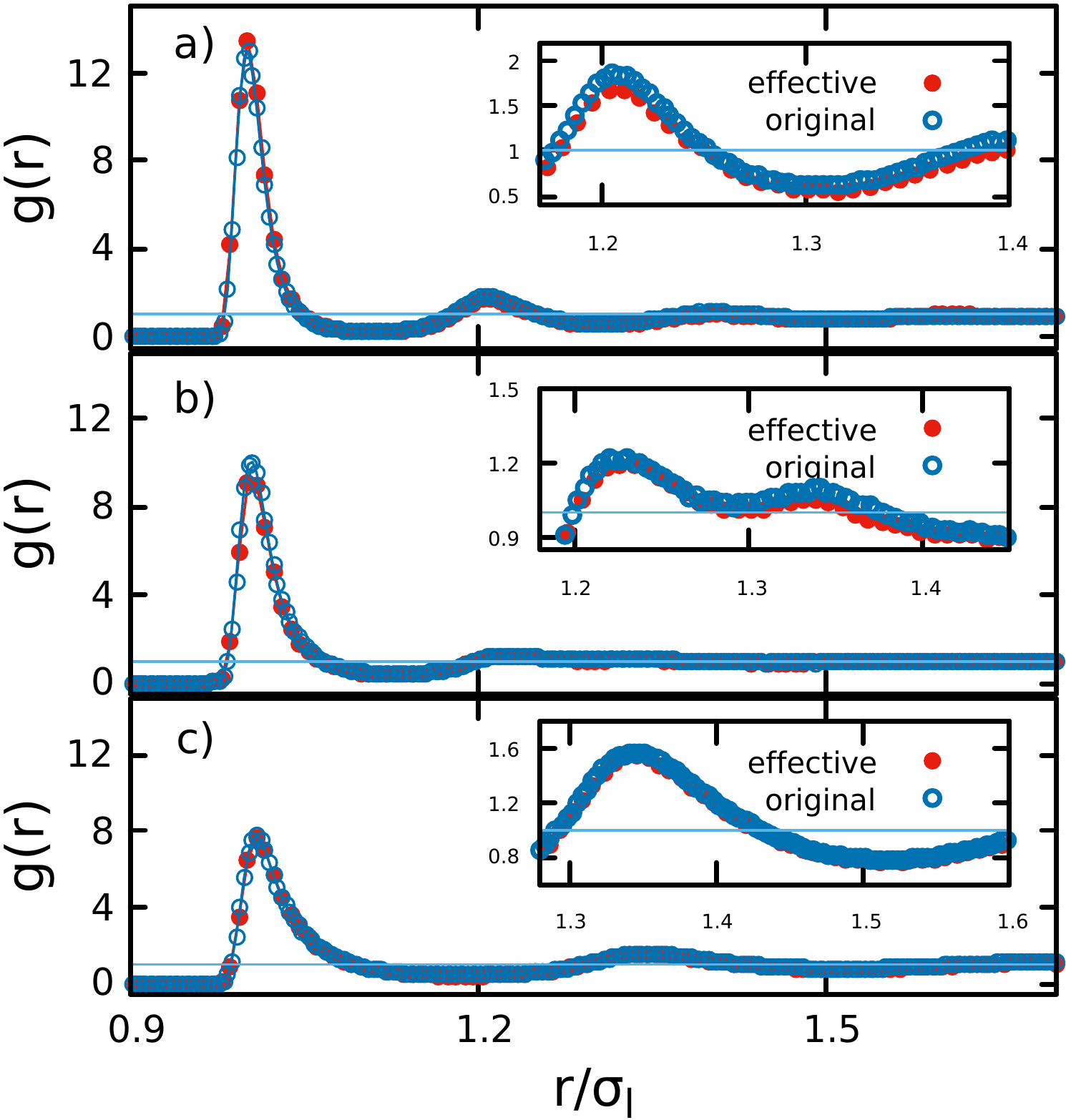}
\caption{Radial distribution functions between large particles for the three systems in Fig.~\ref{fig1}. Insets show closer details around the second minimum. The blue circles are the simulations results for the original mixture with the bare interactions and the red circles the simulations results for the effective monodisperse system of large particles interacting with the depletion forces shown in Fig. \ref{fig1}.}
\label{fig2}
\end{figure}

Figure~\ref{fig2} shows the RDFs between large particles, $g(r)$, for the three systems reported in Fig.~\ref{fig1}. We include a comparison for the original mixture with the bare interactions and simulations results for the effective monodisperse system of large particles interacting with the depletion forces shown in Fig.~\ref{fig1}. Insets show closer details around the second minimum of $g(r)$, where we found the largest deviations. The excellent agreement between RDFs confirms that the contraction procedure, which is carried out by going from Eqs.~(\ref{ftot}) to (\ref{ftotcont}), captures all the depletion effects correctly.

\begin{figure}[htbp!]
\includegraphics[scale=0.55]{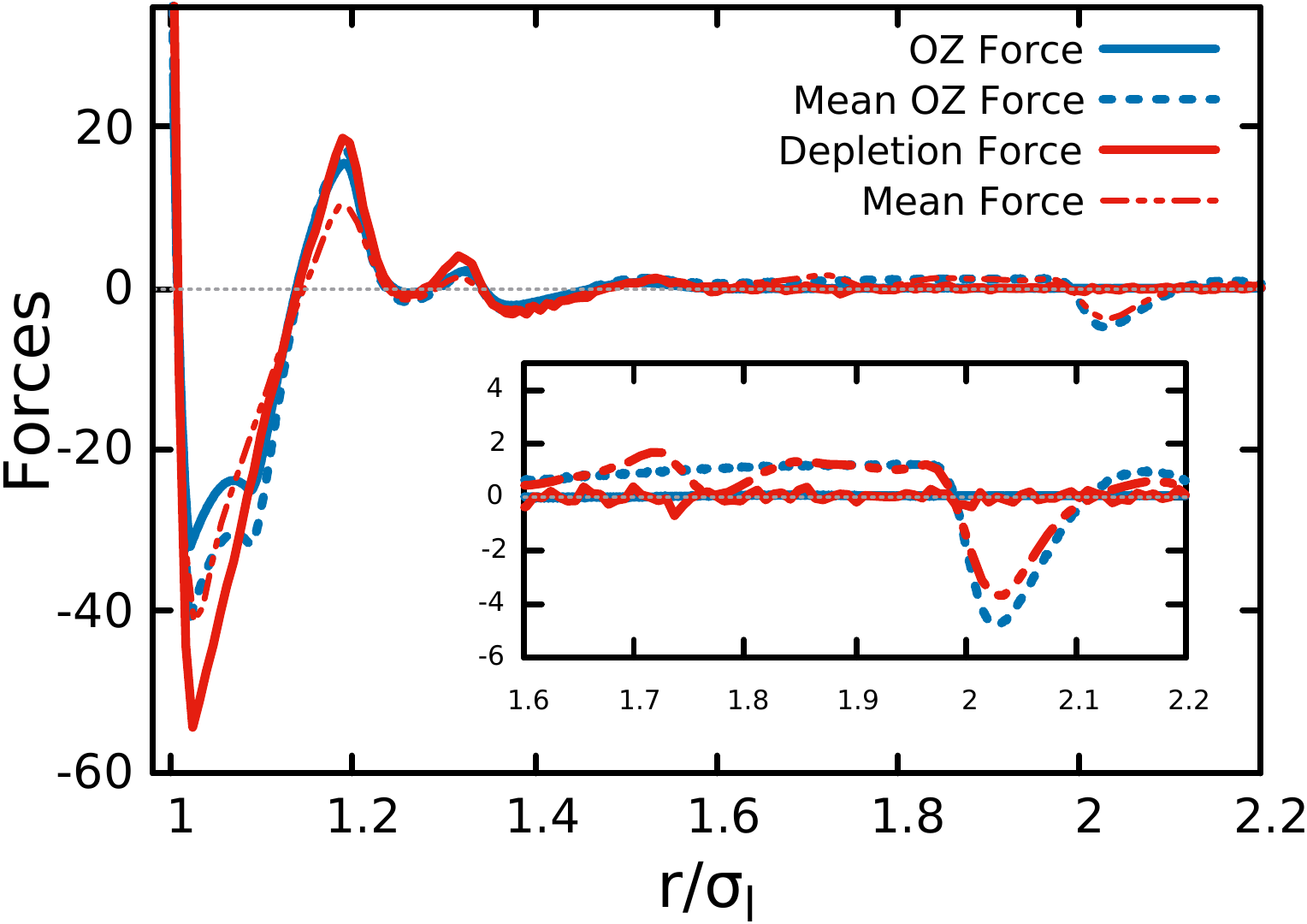}
\caption{Depletion (solid lines) and mean (dashed lines) forces, times $\beta \sigma_{l}$, between large particles for a system with $\phi=0.55$, having $\phi_{l}=0.4$, $\phi_{m}=0.075$, and $\phi_{s}=0.075$. The red lines were obtained using the procedure presented in this letter; the blue lines with the integral equations approach for the depletion forces~\cite{Nestor2021}. Mean OZ force stands for the mean force obtained from the $g(r)$ resulting from the solution of the OZ equation.}
\label{fig3}
\end{figure}

The description is contracted within the integral equations formalism by demanding $g(r)$ to be invariant~\cite{alcaraz-klein,ramon-ro-al-2}. In contrast, the approach presented here requires $\bm{F}^l_i (\bm{r}_i )$ not to change by going from Eqs.~(\ref{ftot}) to (\ref{ftotcont2}). Both conditions are actually equivalent, as suggested by the excellent agreement shown in Fig.~\ref{fig2}. The potential of mean force, $w(r)=-k_B T \ln g(r)$, is defined to reproduce the mean force acting on a large particle, i.e., $\langle \bm{F}^l_i (\bm{r} ) \rangle=-\bm{\nabla} w(r)$~\cite{phillies2000}. Therefore, by demanding the invariance of $\bm{F}^l_i (\bm{r}_i )$ under contractions of the description, we also impose that of $g(r)$. However, $w(r)$ should not be confused with $u^\mathrm{eff}_{ll}(r)$, as it becomes clear from Fig.~\ref{fig3}, since the ensemble average $\langle \bm{F}^l_i (\bm{r}) \rangle$ is over the configurations of large and small particles.

Figure~\ref{fig3} shows the depletion and mean forces, times $\beta \sigma_{l}$, between large particles for a system with $\phi=0.55$, having $\phi_{l}=0.4$, $\phi_{m}=0.075$, and $\phi_{s}=0.0.075$. The mean force was obtained by deriving $w(r)=-\ln g(r)$ numerically using the 3-7 Savitsky-Golay filter~\cite{numerical-recipes}. The discrepancies between both quantities are evident in this concentrated system, mainly at the depletion well and barrier, and around the layer of second neighbors of large particles, $r/\sigma_l \approx 2$, where the mean force shows a structure not included in the depletion one. This is produced by the bare interaction between large particles, not contained in the correlations mediated by small ones, which fade at shorter distances. The potential of mean force is not longer a good approximation for the depletion interactions at those concentrations of large particles. Figure~\ref{fig3} also shows the results obtained using the integral equations approach for the depletion forces~\cite{alcaraz-klein,ramon-ro-al-2} solving the Ornstein-Zernike (OZ) equation with the modified Verlet closure relation~\cite{Nestor2021}. One can see for the first time how that approximation fails at short distances, but is excellent elsewhere. 

There is nevertheless a situation in which depletion forces do not depend on the concentration of the large particles. It happens when the chemical potential of the depletants remains fixed while one adds large particles to the system. The origin of the depletion interaction is the change of the free energy when putting depletants into or taking them from the region in between two large particles~\cite{AO1,lekkerkerker}. Since the free energy does not change by exchanging particles when the chemical potential remains the same, neither does the depletion interaction. We tested this prediction by implementing grand canonical simulation codes in the context of our theoretical scheme and did not observe changes within the numerical error of our calculations (data not shown). Indeed, it constitutes an additional self-consistency test of our proposal.

As in the case of the integral equations theory of depletion forces, the physical approach reported here is much more general than the example used to illustrate it. In principle, both approaches may obtain any effective interaction resulting from reducing the description of the system, be it species, the number of particles, geometry or dimensionality. However, its implementation must be adapted to each situation. In particular, we have shown its outstanding performance in concentrated binary and ternary mixtures of hard-like particles, where entropic depletion is the dominant mechanism behind the interaction between larger particles. Its use to study more sophisticated situations, for example, effective interactions near thermodynamic instabilities, is in progress. Moreover, this approach may also be used in experiments, such as confocal videomicroscopy, to determine the effective forces between colloids.

\begin{acknowledgments}
Authors thank Conacyt for financial support (Grant No. 237425). GP-A thanks Conacyt for financial support (Grant A1-S-46572).
\end{acknowledgments}

\bibliography{PRLDForces.bib}% Produces the bibliography via BibTeX.
\end{document}